\documentclass[preprint,showpacs,preprintnumbers,amsmath,amssymb,natbib]{revtex4}
\usepackage{graphicx}
\usepackage{epsfig}		
\usepackage{dcolumn}
\usepackage{bm}
\usepackage{subfigure}
\def\0{\mbox{\tiny $0$}}
\def\1{\mbox{\tiny $1$}}
\def\2{\mbox{\tiny $2$}}
\def\3{\mbox{\tiny $3$}}
\def\4{\mbox{\tiny $4$}}
\def\5{\mbox{\tiny $5$}}
\def\6{\mbox{\tiny $6$}}
\def\7{\mbox{\tiny $7$}}
\def\8{\mbox{\tiny $8$}}
\def\9{\mbox{\tiny $9$}}

\def\f14{\mbox{\tiny $\frac{1}{4}$}}

\begin{document}

\title{Gaussian fidelity distorted by external fields}

\author{Jonas F. G. Santos and Alex E. Bernardini\footnote{E-mail: jonas@df.ufscar.br, alexeb@ufscar.br}} 
\affiliation{Departamento de F\'isica, Universidade Federal de S\~ao Carlos, PO Box 676, 13565-905, S\~ao Carlos, SP, Brasil.}

\date{\today}

\begin{abstract}
Gaussian state decoherence aspects due to interacting magnetic-like and gravitational fields are quantified through the quantum fidelity and Shannon entropy in the scope of the phase-space representation of elementary quantum systems.
For Gaussian Wigner functions describing harmonic oscillator states, an interacting external field destroys the quantum fidelity and introduces a quantum beating behavior.
Likewise, it introduces harmonic profiles for free particle systems.
Some aspects of quantum decoherence for the quantum harmonic oscillator and for the free particle limit are also quantified through the Shannon entropy.
For the gravitational quantum well, the effect of a magnetic-like field on the quantum fidelity is suppressed by the linear term of the gravitational potential.
To conclude, one identifies a fine formal connection of the quantum decoherence aspects discussed here with the noncommutative quantum mechanics.
\end{abstract}

\maketitle

\section{Introduction}

Gaussian quantum correlations are in the core of the quantum information issues involving continuous variable systems. In addition, Gaussian states are also the elementary blocks in building vacuum states, thermal states and coherent states \cite{FidDef01}, as they play a fundamental role in quantum optics, in low dimensional physics, or even as an effective tool for describing atomic ensembles \cite{Michael}.
Besides providing the necessary theoretical tools for the understanding and the manipulation of quantum correlated systems, the representation of Gaussian states through the Wigner formalism  works as bridge to the classical dynamics.
From a phenomenological point of view, Gaussian Wigner functions can be parametrically manipulated as to describe a set of measured data \cite{Niranjan, Bertet} circumstantially correlated to the issues of quantum-classical transitions of a physical system \cite{Giulim}, for instance, as an indicator of quantum chaos \cite{Kowalewska}. 

The inclusion of external fields into the Hamiltonian that drives the behavior of Gaussian states may bring up typical decoherence and dissipation with recognized phenomenological appeal. 
External fields acting on specific quantum systems are frequently implemented through of quantum simulations, where a kind of controllable quantum system is used to study another less accessible one \cite{Georgescu}. 
As a typical example, the effect of a constant magnetic field on atoms has revealed the split in the energy spectrum, in the well-known phenomenon of Zeeman effect \cite{Robinett}. 

Given that the Wigner function in the phase-space quantum mechanics is connected with the information which can be obtained from a quantum system, the quantum fidelity, computed through a Gaussian envelop, can encompass the decoherence aspects of the dynamical evolution of such elementary quantum systems.
In this letter, the influence of magnetic external fields on the quantum fidelity of two dimensional harmonic oscillators, their corresponding free particle limit, and an extension as to include the gravitational quantum well (GQW) dynamics, are therefore quantified through the Gaussian Wigner formalism.

Our manuscript is organized as follows.
In section {II} the Wigner-Weyl formalism of the quantum mechanics, as well as the way to obtain the quantum fidelity and the Shannon entropy are introduced. The Gaussian Wigner state to be used throughout the subsequent analysis of some particular quantum systems is presented. In section {III}, one reports about the Wigner phase-space formalism for the harmonic oscillator in the presence of an external interacting magnetic field. The classical limit for the free particle system is obtained by setting null the natural frequency of the oscillator, $\omega_0 = 0$. The Gaussian state fidelity and a qualitative analysis for the Shannon entropy are also evaluated. In section IV, the quantum fidelity and the quantum decoherence aspects for the gravitational quantum well are discussed in the phase-space framework.
A fine formal connection between external magnetic field interacting systems and the noncommutative quantum mechanics is noticed in section V.
Our conclusions are drawn in section VI.

\section{The Weyl-Wigner formalism of the quantum mechanics, quantum fidelity, and Shannon entropy}

By identifying the density matrix of a quantum system with $\hat{\rho} = |\Psi \rangle \langle \Psi |$, one can define a Wigner function through the Weyl transform as \cite{Case,Wigner},
\begin{equation}
W(r, p) =  h^{-1} \rho^W = 
\int \hspace{-.15cm}ds\,\exp{\left[i \, p \,s/\hbar\right]}\,
\Psi(r - s/2)\,\Psi^{\ast}(r + s/2),
\label{Wignerfunction}
\end{equation}
which can be naturally generalized to a statistical mixture, such that the expectation value of an observable $\hat{O}$ can be computed through 
\begin{equation}
\langle O \rangle = 
\int \hspace{-.15cm}\int \hspace{-.15cm} dr\,dp \,W(r, p)\,{O^W}(r, p).
\end{equation}

The probability distributions for $r$ and $p$ are equivalently given by
\begin{equation}
\int \hspace{-.15cm} dr\,W(r, p) = \Phi^{\ast}(p)\,\Phi(p)
\quad\mbox{and}\quad
\int \hspace{-.15cm} dp \,W(r, p)= \Psi(r)^{\ast}\Psi(r),
\label{dist1}
\end{equation}
such that the Wigner function can also be computed from $\Phi(p)$ through
\begin{equation}
W(r, p) = 
\int \hspace{-.15cm}ds\,\exp{\left[-i\, r \,s/\hbar\right]}\,
\Phi(p - s/2)\Phi^{\ast}(p + s/2),
\end{equation}
Additional properties related to the density matrix theory can be obtained from the above prescription \cite{Hillery,Lee}. For instance, the Weyl transform of an operator has intrinsic properties that allow one to write the quantum fidelity, $F$, in terms of Wigner functions.
The quantum fidelity, $F$, is a commonly used measure to compare an input state and an output state through a given quantum channel \cite{FidDef01,FidTPT, FidChaos}, as it works as a kind of decoherence quantifier. If $F$ goes to unity, it means that the output state is very similar to the input state. Likewise, if $F$ goes to zero, the output is completely different from the input state. 
Effectively, the fidelity measures the projection, varying from zero to unity, of a time-evolving state onto a departure state.

By using the Weyl transform of an operator and the property of the trace of the product of two operators, one has
\begin{equation}
Tr [\hat{A} \hat{B}] = \int \int A^W(r, p)\,B^W(r, p)\,dr dp,
\label{prop1}
\end{equation}
which can be used into the definition of the quantum fidelity \cite{FidDef01},
\begin{equation}
F (\hat{\rho}_1, \hat{\rho}_2) = \left[Tr(\sqrt{\sqrt{\hat{\rho}_1}\hat{\rho}_2\sqrt{\hat{\rho}_1}})\right]^2,
\end{equation}
where $\hat{\rho}_1$ and $\hat{\rho}_2$ are two states of the quantum system.
Noticing that
\begin{equation}
Tr[(\hat{\rho}_1 \hat{\rho}_2)^{1/2}] = \int \int (\rho_1^W \rho_2^W)^{1/2} \,dr dp,
\end{equation}
and using the Wigner function from Eq.~(\ref{Wignerfunction}), one obtains
\begin{equation}
Tr[(\hat{\rho}_1 \hat{\rho}_2)^{1/2}] = \int \int (W_1 W_2)^{1/2} \,dr dp,
\end{equation}
which, through the definition of $F$, leads to
\begin{equation}
F = \left[\int \int (W_1 W_2)^{1/2} \,dr dp\right]^2,
\end{equation}
which can be re-written in terms of the coordinates of a $2$-mode phase-space as
\begin{equation}
F = \left[\int \int \int \int(W_1(x, p_x, y, p_y) W_2(x, p_x, y, p_y))^{1/2} dx\, dp_x\, dy\, dp_y\right]^2.
\label{fidelity}
\end{equation}

To investigate the decoherence effect due to external fields, one assumes $W_1$ as being the initial state, i. e. the quantum state before the external field be turned on, then one inserts the equations of motion corresponding to each system, and finally, one measures how far away the external field-dependent state, $W_2$, will be from $W_1$. 
Throughout this work, it will be considered the following normalized initial Gaussian Wigner function,
\begin{equation}
W(\mbox{\bf r},\mathbf{p};t) = \frac{1}{\pi^2}\exp\left[-\left({x}(t) - x_0)^2 + ({y}(t) - y_0)^2\right)\right]\exp\left[-\left({p}_x(t) - p_{x_0})^2 + ({p}_y(t) - p_{y_0})^2\right)\right].
\label{gaussian1}
\end{equation}
where one identifies the canonical coordinates by $\mbox{\bf r}\sim \{x,y\}$ and $\mbox{\bf p}\sim \{p_x,p_y\}$.

Likewise, obtaining the decoherence information and $2$-mode quantum correlations of a quantum system can also be performed through quantifying the von Neumann entropy, $S(\hat{\rho}) = -\mbox{Tr}[\hat{\rho}\, \mbox{ln}(\hat{\rho})]$, where $\hat{\rho}$ is a quantum state of the system. The equivalent Shannon entropy in the phase-space representation, using the Wigner function, is given by \cite{Adesso01, Parvin01},
\begin{equation}
S_W = - \int \int |W(r,\, p)| \mbox{ln}[|W(r,\, p)]\, dr\, dp,
\label{Entropy}
\end{equation}
which is valid for both Gaussian and non-Gaussian states. 
If one has a two dimensional ($2D$) system where the Wigner function can be separated in the form of
\begin{equation}
W(x, p_x, y, p_y) = W(x, p_x)\, W(y, p_y) = W_x\, W_y, 
\end{equation}
it is easy to show that the Shannon entropy can be written as
\begin{equation}
S_W = S_{W_x} + S_{W_y},
\end{equation}
that is, there is no information shared between the two sub-systems and consequently the mutual information is zero. However, when one has an external interacting field acting on the system, it is necessary to calculate the Shannon entropy of the system as a whole.

\section{$2D$-harmonic oscillator coupled to magnetic fields}

Let one considers the case of a $2D$ harmonic oscillator of mass $m$, frequency $\omega_0$ and charge $q$ in a magnetic field $B_0$ perpendicular to the $2D$ plane.
The Hamiltonian in this case is written as \cite{Ballentine,Bernardini01,Rosenbaum}, 
\begin{equation}
{H}(\textbf{r}, \textbf{p}) =  \frac{\textbf{p}^2}{2m} - \frac{q B_0}{2m}{L}_z + \frac{q^2 B_0^2}{8m}({x}^2 + {y}^2
)+ \frac{m \omega_0^2}{2}({x}^2 + {y}^2),
\end{equation}
where $B_0$ is the intensity of the external magnetic field and ${L}_z$ is the angular momentum in the $z-$direction, perpendicular to the harmonic motion. 

By defining $\omega = q B_0/2m$ as being the frequency of coupling, and defining the new parameters \cite{Bernardini01,Rosenbaum}
\begin{equation}
\lambda^2 = \frac{m}{2}(\omega^2 + \omega^2_0), \quad \kappa^2 = \frac{1}{2m},
\end{equation}
the Weyl transform of the Hamiltonian reads,
\begin{equation}
H^W(r_i, p_i) = \lambda^2 r_i^2 + \kappa^2 p_i^2 + \omega \epsilon_{ij} p_i r_j.
\label{HamilHO}
\end{equation}

Given that, from Eq.~(\ref{HamilHO}), the variables $r_i$ and $p_i$ satisfy the Hamilton equations of motion, one obtains the following set of coupled first-order differential equations \cite{Bernardini01},
\begin{eqnarray}
\dot{p}_k &=& -\frac{i}{\hbar} \left[p_k,\,H^W\right] = -2 \lambda^2\,\mathit{r}_k - \omega\,\varepsilon_{jk}p_j,\nonumber\\
\dot{\mathit{r}}_k &=& -\frac{i}{\hbar} \left[\mathit{r}_k,\,H^W\right] =  ~~2 \kappa^2\,p_k - \omega\,\varepsilon_{jk}\mathit{r}_j,\qquad k,j = 1,2,
\label{eqs01}
\end{eqnarray}
so that $\mbox{\bf \em q}$ and $\mathbf{p}$ may be interpreted as classical dynamical variables within the Wigner-Weyl formalism.
The above equations can be rewritten as two uncoupled forth-order differential equations as \cite{Bernardini01}
\begin{eqnarray}
\ddddot{p}_k &=& -2(\omega^2 + 4\lambda^2\kappa^2)\ddot{p}_k + (\omega^2 - 4\lambda^2\kappa^2)p_k,\nonumber\\
\ddddot{\mathit{r}}_k &=& -2(\omega^2 + 4\lambda^2\kappa^2)\ddot{q}_k + (\omega^2 - 4\lambda^2\kappa^2)r_k,
\end{eqnarray}
from which one gets the solutions,
\begin{eqnarray}
\mathit{r}_1(t)&=& x_0\cos(\Omega t)\cos(\omega t) + y_0\cos(\Omega t)\sin(\omega t) + \frac{\kappa}{\lambda}\left[p_{y_0}\sin(\Omega t)\sin(\omega t) + p_{x_0}\sin(\Omega t)\cos(\omega t)\right],\nonumber\\
\mathit{r}_2(t)&=& y_0\cos(\Omega t)\cos(\omega t) - x_0\cos(\Omega t)\sin(\omega t) - \frac{\kappa}{\lambda}\left[p_{x_0}\sin(\Omega t)\sin(\omega t) - p_{y_0}\sin(\Omega t)\cos(\omega t)\right],\nonumber\\
p_1(t)&=& p_{x_0}\cos(\Omega t)\cos(\omega t) + p_{y_0}\cos(\Omega t)\sin(\omega t) - \frac{\lambda}{\kappa}\left[y_0\sin(\Omega t)\sin(\omega t) + x_0\sin(\Omega t)\cos(\omega t)\right],\nonumber\\
p_2(t)&=& p_{y_0}\cos(\Omega t)\cos(\omega t) - p_{x_0}\cos(\Omega t)\sin(\omega t)+ \frac{\lambda}{\kappa}\left[x_0\sin(\Omega t)\sin(\omega t) - y_0\sin(\Omega t)\cos(\omega t)\right],\qquad
\label{solutionsHO}
\end{eqnarray}
where $x_0$, $y_0$, $p_{x_0}$ and $p_{y_0}$ are arbitrary boundary initial values and
\begin{equation}
\Omega = 2\lambda \kappa = \left(\omega^2 + \omega_0^2\right)^{1/2}.
\end{equation}

Following the Wigner-Weyl formalism of the quantum mechanics, the {\em star}genfunctions for the Hamiltonian of a harmonic oscillator \cite{Bernardini01,Rosenbaum} in a magnetic field are obtained from the {\em star}genvalue equation \cite{Bernardini01,Rosenbaum},
\begin{equation}
W_{n_1, n_2}(\textbf{r}, \textbf{p}) = \frac{(-1)^{n_1 +n_2}}{\pi^2 \hbar^2}\mbox{exp}\left[-\frac{1}{\hbar}\left(\frac{\lambda}{\kappa}\textbf{r}^2 + \frac{\kappa}{\lambda}\textbf{p}^2\right)\right] L_{n_1}^0(\Omega_+/\hbar)L_{n_2}^0(\Omega_-/\hbar),
\label{WignerHO}
\end{equation}
where $L_n^0$ are the associated Laguerre polynomials, $n_1$ and $n_2$ are non-negative integers, and
\begin{equation}
\Omega_\pm = \frac{\lambda}{\kappa}\textbf{r}^2 + \frac{\kappa}{\lambda}\textbf{p}^2 \mp 2\Sigma_{i,j = 1}^2(\epsilon_{ij} r_i p_j),
\end{equation}
such that the energy spectrum is given by \cite{Bernardini01,Rosenbaum}
\begin{equation}
E_{n_1, n_2} = \hbar[2\lambda \kappa(n_1 + n_2 + 1) + \omega(n_1 - n_2)].
\end{equation}

The Wigner function for the harmonic oscillator in an external magnetic field in Eq.~(\ref{WignerHO}) is a stationary function. Gaussian Wigner functions may be used to compute quantum fidelity as to investigate the influence of external fields on the free harmonic oscillator. By inserting the equations of motion from Eq.~(\ref{solutionsHO}) into the Gaussian Wigner function from Eq.~(\ref{gaussian1}), one can quantify the fidelity between an initial Gaussian state which evolves in time as a state coupled to an external magnetic field.

\subsection{The limit case when $\omega_0$ = 0}

When the frequency of the harmonic oscillator, $\omega_0$, goes to zero, the system above is reduced to a free particle moving in a magnetic field \cite{Ballentine,2015}. In this case one has the Hamiltonian of the Zeeman effect \cite{Robinett}.
After some simple mathematical manipulations, setting $\omega_0 = 0$ into Eq.~(\ref{solutionsHO}), one obtains the following solutions for the equations of motion \cite{2015},
\begin{eqnarray}
\mathit{r}_1(t)&=& \frac{1}{2}\left[\left(x_0 + \frac{p_{y_0}}{m\omega}\right) + \left(x_0 - \frac{p_{y_0}}{m\omega}\right)\cos(2\omega t) + \left(y_0 + \frac{p_{x_0}}{m\omega}\right)\sin(2\omega t)\right],\nonumber\\
\mathit{r}_2(t)&=& \frac{1}{2}\left[\left(y_0 - \frac{p_{x_0}}{m\omega}\right) + \left(y_0 + \frac{p_{x_0}}{m\omega}\right)\cos(2\omega t) - \left(x_0 - \frac{p_{y_0}}{m\omega}\right)\sin(2\omega t)\right],\nonumber\\
p_1(t)&=& \frac{1}{2}\left[\left(p_{x_0} - m\omega\, y_0\right) + \left(p_{x_0} + m\omega\, y_0\right)\cos(2\omega t) + \left(p_{y_0} - m\omega\, x_0\right)\sin(2\omega t)\right],\nonumber\\
p_2(t)&=&\frac{1}{2}\left[\left(p_{y_0} + m\omega\, x_0\right) + \left(p_{y_0} - m\omega\, x_0\right)\cos(2\omega t) - \left(p_{x_0} + m\omega\, y_0\right)\sin(2\omega t)\right],
\label{solutionsFree}
\end{eqnarray}
and the stationary Wigner function for this system turns into \cite{2015}
\begin{equation}
W_{n}(\mbox{\bf  r},\mathbf{p}) = \mathcal{N}\frac{(-1)^n}{\pi \hbar}\mbox{exp}\left[-\Omega/\hbar \right]\, L_n^{0}\left(\Omega/\hbar\right),
\label{WignerFP}
\end{equation}
where
\begin{equation}
{\Omega}(t)= {\lambda\over\kappa}\mbox{\bf r}^2(t) + {\kappa\over\lambda}\mathbf{p}^2(t) + 2 \sum_{i,j = 1}^2{\left(\epsilon_{ij}p_i(t) \mathit{r}_j(t)\right)},
\end{equation}
and now one has the constraint
\begin{equation}
2\lambda \kappa = \omega,
\end{equation}
which allows one to write the energy spectrum as \cite{2015},
\begin{equation}
E_n = \hbar \omega (2n + 1).
\end{equation}

By performing the integral in Eq.~(\ref{fidelity}) for the $2D$ harmonic oscillator coupled to magnetic fields, one obtains the following analytical expression for the fidelity,
\begin{equation}
F(\omega, t) = \exp\left[(p_{x_0}^2 + p_{y_0}^2 + x_0^2 + y_0^2)(-1 + \cos(t)  \cos(\omega t)) + 2(p_{y_0} x - p_{x_0} y)\sin(t)\sin(\omega t)\right],
\end{equation}
where one has assumed $\omega_0 = 1$.

\begin{figure}
\hspace{-1.5 cm}
\centering
\includegraphics[scale=0.80]{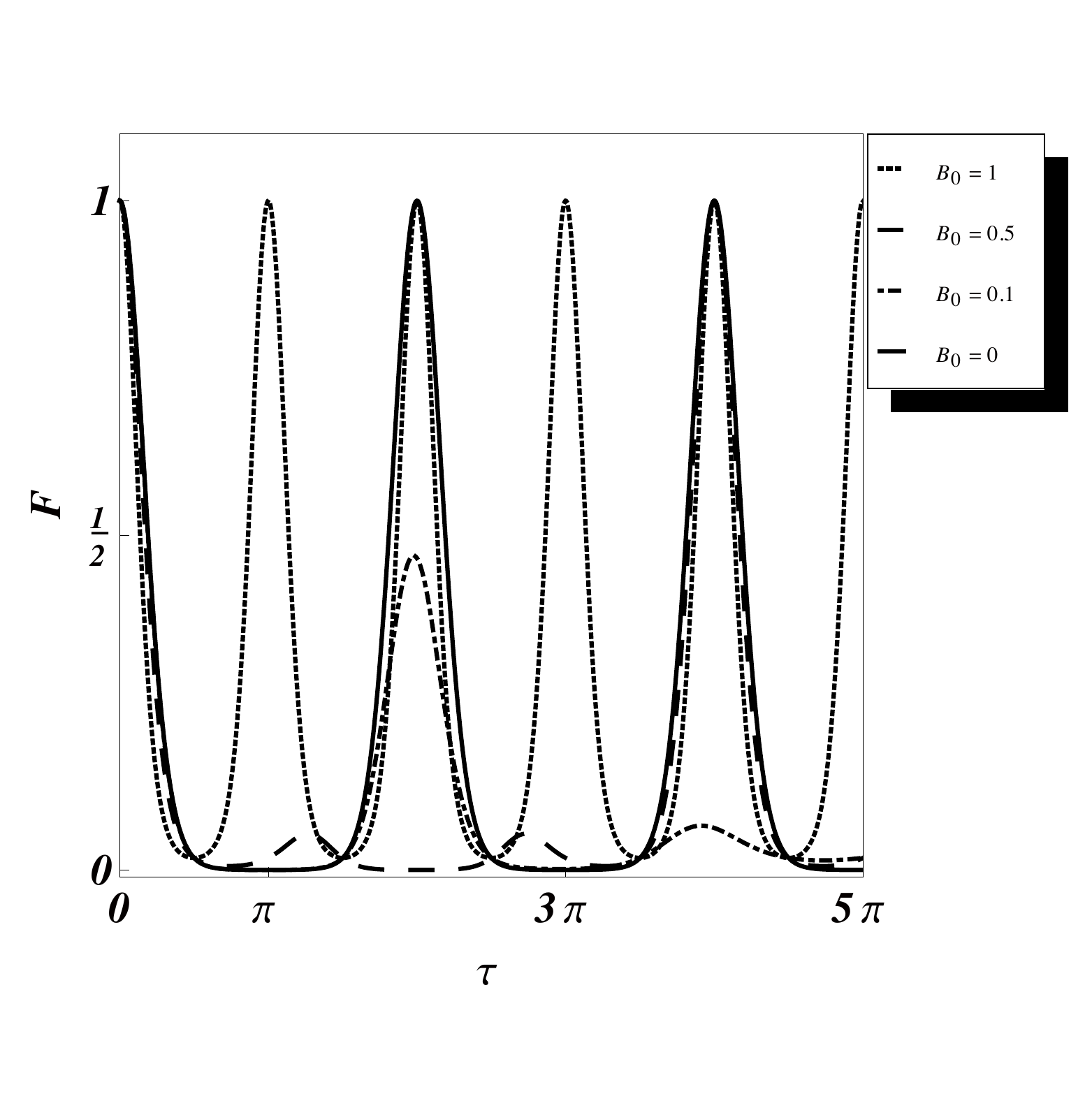}
\caption{\small Gaussian quantum fidelity as function of time, $\tau$, for a quantum state evolving according to the harmonic oscillator dynamics.
One has considered $m = \hbar = 1$, $g = 2$ and $x =  y = p_{x_0} = p_{y_0} = 1$, with $B_0 = 0.1$, $0.5$ and $1$.
For increasing values of the magnetic field, one has increasing values for the frequencies, $\omega$, of the beating behavior.}
\label{FidHONC}
\end{figure}

\begin{figure}
\hspace{-1.5 cm}
\centering
\includegraphics[scale=0.8]{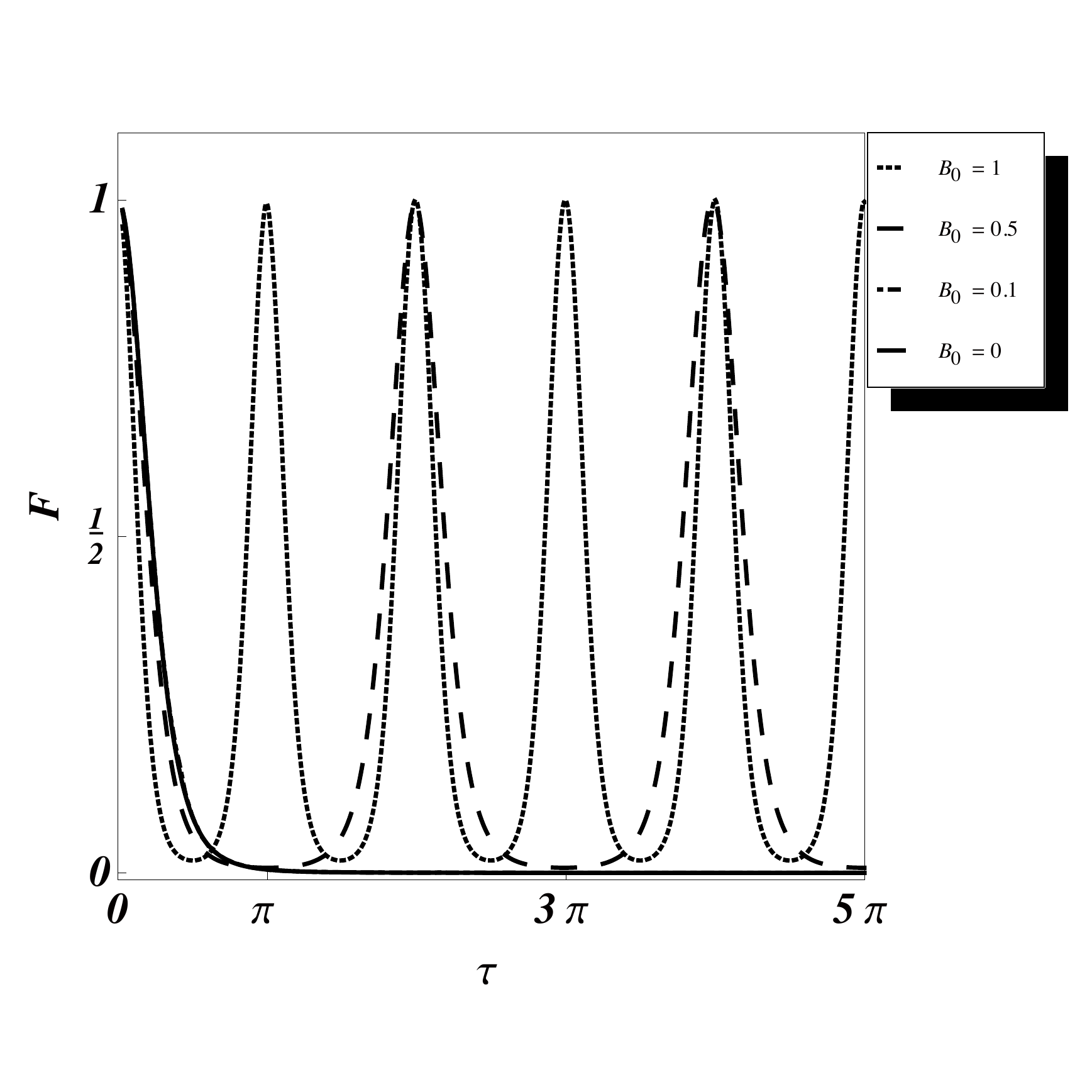}
\caption{\small Gaussian quantum fidelity as function of time, $\tau$, for a quantum state evolving according to the free particle dynamics.
One has considered $m = \hbar = 1$, $g = 2$ and $x =  y = p_{x_0} = p_{y_0} = 1$, with $B_0 = 0.1$, $0.5$ and $1$.
For increasing values of the magnetic field, the harmonic behavior is quickly recovered.}
\label{FidFreeNC}
\end{figure}

Figs. \ref{FidHONC} and \ref{FidFreeNC} depict the Gaussian fidelity for the harmonic oscillator and for the free particle system, respectively, for three cases, $\omega = 0.1$, $0.5$, $1$ and for the case when there is no magnetic field, $\omega = 0$. 

\begin{figure}
\hspace{-1.5 cm}
\centering
\includegraphics[scale=0.9,bb = 0 0 350 480]{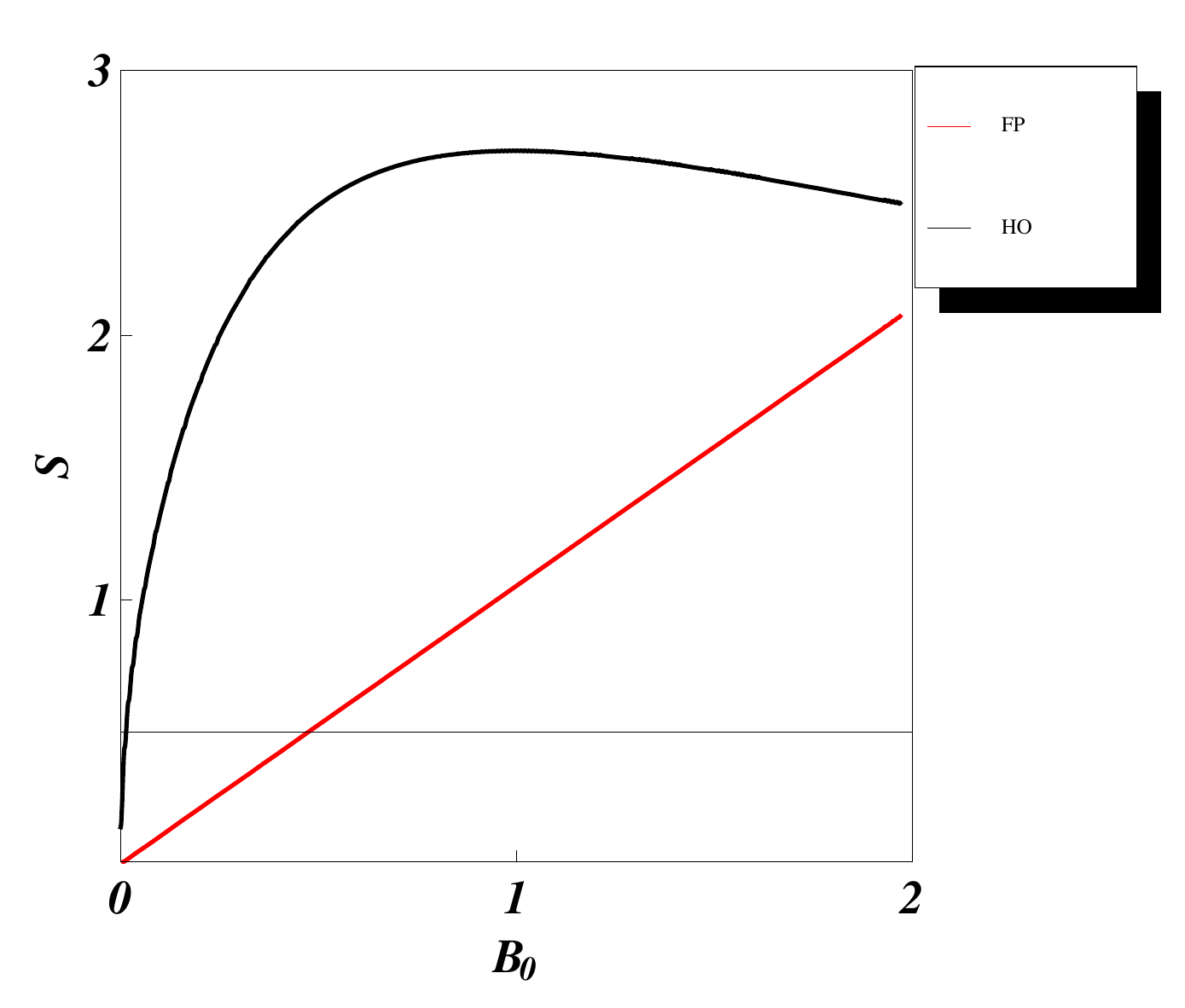}
\caption{\small Shannon entropy for the ground states of the harmonic oscillator (black line) and free particle (red line) systems. As the intensity of the magnetic field $B_0$  goes to zero the Shannon entropy also goes to zero.}
\label{Shannonentro}
\end{figure}

From Figs. \ref{FidHONC} and \ref{FidFreeNC}, one notices that for $\omega \neq 0$ the Gaussian fidelity assumes a periodic characteristic as function of the frequency of the external field. 
From the point of view of the Shannon entropy, one can obtain some insight about the influence of the magnetic field on the information of the state in the phase-space. 
In order to illustrate the Shannon entropy for the harmonic oscillator and for the free particle system subjected to an external field, one considers the stationary Wigner functions (\ref{WignerHO}) and (\ref{WignerFP}) and obtains the Shannon entropy as function of the intensity of the magnetic field. The result is depicted in Figs. \ref{Shannonentro} for the ground state of each system.

Harmonic oscillator and free particle induced ground states by the presence of an external magnetic field are discussed in terms of phase-space Gaussian profiles. A Gaussian highly localized means that one has considerable information about the state of the system. Therefore, the effect of $B_0$ on the Shannon entropy increases the localization of the Gaussian states.
When $B_0$ vanishes, the Shannon entropy also vanishes, which means that one has a very spread-out Gaussian. On the other hand, as the value of $B_0$ increases, the states of the systems turns to be more localized, which reflects into an increasing of the the Shannon entropy.

For the harmonic oscillator, there exist a maximum information which one can access, due to dimension of the oscillator. This fact is represented in the Fig. \ref{Shannonentro} for the maximum value of entropy. The same does not happens to the free particle system, once it is always possible to restrict the localization of a particle through, for instance, a delta function. Fig. \ref{Shannonentro} shows that the Shannon entropy goes to infinity when $B_0$ goes to infinity, that is, one needs an extremely large amount of information to have a particle completely localized. When $B_0$ vanishes, the particle can be found anywhere.

\section{The gravitational quantum well}

The gravitational quantum well (GQW) for ultra cold neutrons (UCN) are supposed to be a challenging testable platform of quantum mechanics, since it should allow the detection of quantum bound states in a classical potential, i. e. due to the gravitational coupling between the Earth and ultra-cold neutrons \cite{Nesvizhevsky,Ichikawa,Baefler,Yoshiki, Kitaguchi,Nico}.

The Hamiltonian of a particle in a gravitational potential is given by
\begin{equation}
{H}(\textbf{r}, \textbf{p}) = \frac{1}{2m}\textbf{p}^2 +  m g y,
\label{HamLin}
\end{equation}
from which equations of motion lead to a set of solutions given by
\begin{eqnarray}
x(t) &=& x_0 + \frac{p_{x0}}{m}t ,\nonumber\\
y(t) &=& y_0 - \frac{p_{y0}}{m}t - \frac{g}{2}t^2,\nonumber\\
p_x(t) &=& p_{x0}, \nonumber\\
p_y(t) &=& p_{y0} - m g t.
\label{EMGQW}
\end{eqnarray}

The Wigner function, {\em star}genfunction of ${H}$, can be obtained by separating the Hamiltonian into two peaces respectively related to the $x$-coordinate, which exhibits a free particle profile, and to the $y$-coordinate, such that,
\begin{equation}
H^W_y = \frac{p^2_y}{2m} + m g y,
\end{equation}
and the {\em star}genvalue equation reads
\begin{equation}
H_y^w \star W_y = E_y.
\end{equation}

By operating the above introduced Moyal $\star$ product \cite{Moyal1,Moyal2}, one obtains the following differential equation,
\begin{equation}
\left[\xi - \frac{\hbar^2 m g^2}{8}\frac{\partial^2}{\partial \xi^2} - E_y\right] W_y(\xi) = 0,
\end{equation}
where
\begin{equation}
\xi = \frac{p_y^2}{2m} + m g y.
\end{equation}

The solution for $W_y(\xi)$ is an Airy function, $Ai(\xi)$, given by
\begin{equation}
W_y(y, p_y) = A_n\, Ai\left[\left(\frac{8}{m g^2 \hbar^2}\right)^{1/3}\left(\frac{p_y^2}{2m} + m g y - E_{n_y}\right)\right],
\end{equation}
where the energy spectrum is determined by the zeros of the wave function as
\begin{equation}
E_{n_y} = - \left(\frac{m g^2 \hbar^2}{2}\right)^{1/3} \lambda_{n_y},
\end{equation}
where $\lambda_{n_y}$ are the roots of the Airy function, and the unitarity of the Wigner function guarantees that the normalization factor is given by
\begin{equation}
A_n = \left[\int_0^\infty \int_{-\infty}^\infty Ai\left[\left(\frac{8}{m g^2 \hbar^2}\right)^{1/3}\left( \frac{p_y^2}{2m} + m g y - E_{n_y}\right)\right]\right]^{-1}.
\end{equation}

For the free particle $x$-coordinate driven by
\begin{equation}
H^W_x = \frac{1}{2m}p_x^2,
\end{equation}
the localization conditions are assumed to constrain the Wigner function to Gaussian boundary values as to have the $2$-mode Wigner function given by
\begin{eqnarray}
W(x, p_x, y, p_y;t) &=&  W(x, p_x; t) W(y, p_y; t)\nonumber\\
&=& A_n\,\mbox{G}(x, p_x; t)\, Ai\left[\left(\frac{8}{m g^2 \hbar^2}\right)^{1/3}\left(\frac{p_y^2}{2m} + m g y - E_{n_y}\right)\right],
\label{SolutionGQW}
\end{eqnarray}
where $\mbox{G}(x, p_x; t)$ describes a Gaussian function in the phase-space, and
$W(x, p_x, y, p_y;t)$ exhibits a stationary behavior under the substitution of solutions from Eqs.~(\ref{EMGQW}).

The role played by the gravity in quantum schemes open a special window of possibilities for measuring distortions of quantum mechanics in high energy levels, even when the gravitational well is the responsible for the quantum states measured in the lab. 
For instance, decoherence aspects can be detected through distortions over Gaussian fidelity for quantum states subjected to external (magnetic) field perturbations.
In a very suitable context, the GQW system has also been considered to estimate noncommutative corrections to the standard quantum mechanics \cite{Bertolami,Banerjee}.

A way to perform this is to write the Hamiltonian of the GQW in the presence of an external magnetic field, $B=B_0{z}$, as  
\begin{equation}
H^W_{GQW}(r_i, p_i) =  \lambda^2 r_i^2 + \kappa^2 p_i^2 +  \omega (p_1 r_2 - p_2 r_1) +m\, g\,  r_2,
\label{hamiltNCLin}
\end{equation}
with $\lambda$, $\kappa$ and $\omega$ are the same parameters introduced in the free particle scheme, and one identifies the canonical coordinates by $\{r_1,r_2\}\sim \{x,y\}$ and $\{p_1,p_2\}\sim \{p_x,p_y\}$.

The Hamiltonian from Eq.~(\ref{hamiltNCLin}) leads to two uncoupled third-order differential equations which sets the dynamics of the GQW  distorted by $B_0{z}$ as
\begin{eqnarray}
\dddot{p}_k + 4\,\omega^2\dot{p}_k + 2 m\, g\, \omega^2\, \delta_{1k} &=& 0,\nonumber\\
\dddot{r}_k + 4\,\omega^2 \dot{r}_k - 2 g\, \omega\, \delta_{2k} &=& 0,
\end{eqnarray}
from which one gets the solutions,
\begin{eqnarray}
\mathit{r}_1(t)&=& \frac{1}{2}\left[\left(x_0 + \frac{p_{y_0}}{m\omega}\right) + \left(x_0 - \frac{p_{y_0}}{m \omega}\right)\cos(2\omega t) + \left(y_0 + \frac{p_{x_0}}{m\omega}\right)\sin(2\omega t)\right],\nonumber\\
\mathit{r}_2(t)&=& \frac{1}{2}\left[\left(y_0 - \frac{p_{x_0}}{m\omega}\right) + \left(y_0 + \frac{p_{x_0}}{m\omega}\right)\cos(2\omega t) - \left(x_0 - \frac{p_{y_0}}{m\omega}\right)\sin(2\omega t)\right] + \frac{g}{2\omega}t,\nonumber\\
p_1(t)&=& \frac{1}{2}\left[\left(p_{x_0} - m\omega\, y_0\right) + \left(p_{x_0} + m\omega\, y_0\right)\cos(2\omega t) + \left(p_{y_0} - m\omega\, x_0\right)\sin(2\omega t)\right] + \frac{m g}{2}t,\nonumber\\
p_2(t)&=&\frac{1}{2}\left[\left(p_{y_0} + m\omega\, x_0\right) + \left(p_{y_0} - m\omega\, x_0\right)\cos(2\omega t) - \left(p_{x_0} + m\omega\, y_0\right)\sin(2\omega t)\right].
\label{solutionsLinear}
\end{eqnarray}

\begin{figure}
\hspace{-1.5 cm}
\centering
\includegraphics[scale=0.9,bb = 0 0 350 480]{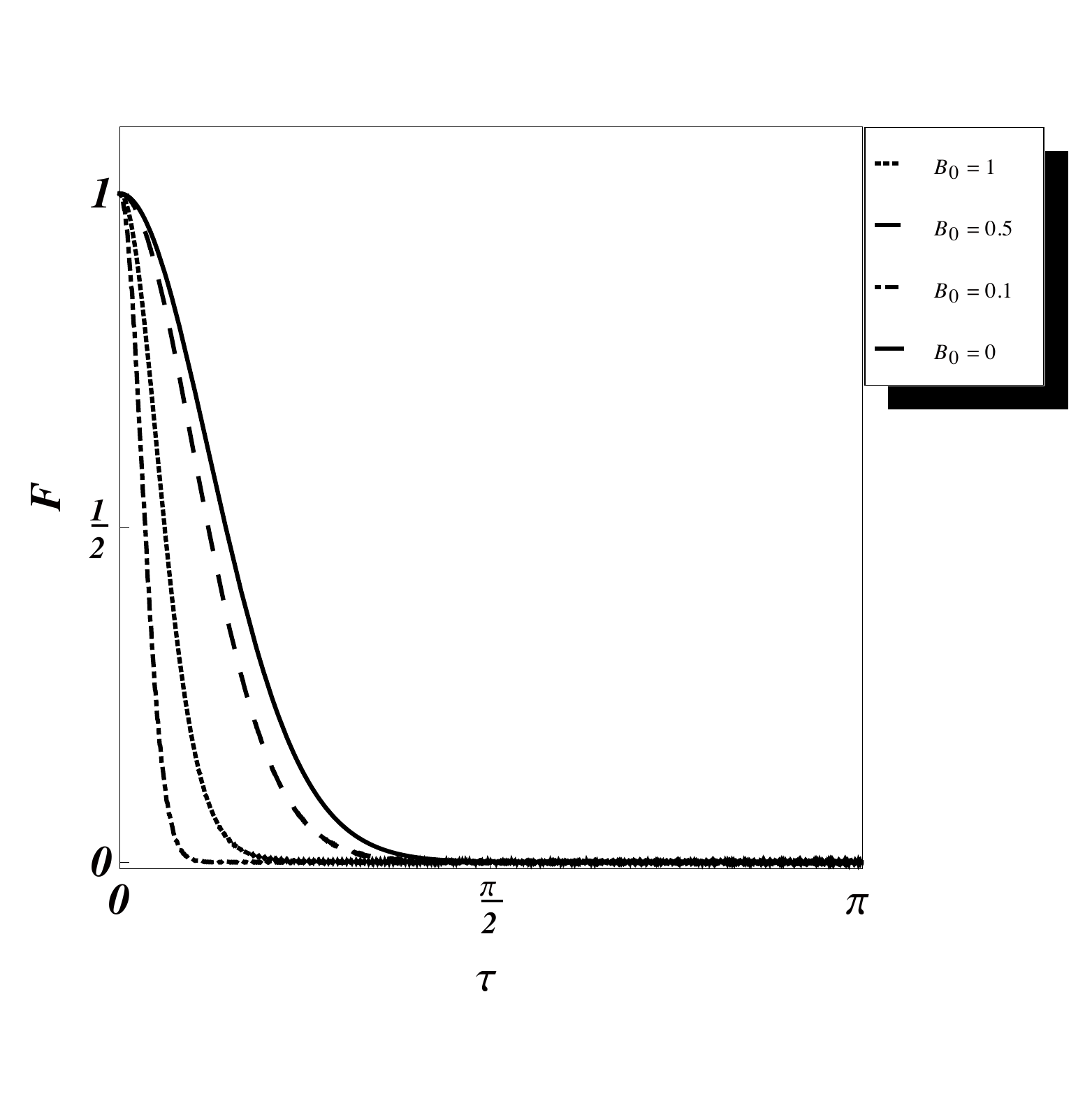}
\caption{\small Gaussian quantum fidelity as function of time, $\tau$, for a quantum state evolving according to the GQW dynamics.
One has considered $m = \hbar = 1$, $g = 2$ and $x =  y = p_{x_0} = p_{y_0} = 1$, with $B_0 = 0.1$, $0.5$ and $1$.
The limiting condition given by $B_0 = 0$ is shown by the dashed line.
For increasing values of the magnetic field, one has an increasing suppression of the Gaussian fidelity.}
\label{FidGQWNC1}
\end{figure}

Fig. \ref{FidGQWNC1} depicts the quantum fidelity for the GQW system in the presence of a uniform magnetic field. As expected, one still has a quick decrease of the graphics, although the inclusion of an external field damps the fidelity faster.

\section{Mapping noncommutative effects through an external magnetic field}

The noncommutative formulation of quantum mechanics \cite{Gamboa}, once it is extended to the phase space, has supported the understanding of coupling and decoherence aspects exhibited by $2D$ quantum oscillators \cite{Rosenbaum,Bernardini01}, the quantum corrections for the ultra-cold neutron energy spectrum in the scope of the GQW problem \cite{Bertolami,07A}, and several other issues\cite{Nekrasov01,Catarina,Bernardini13B,Bernardini13C,Bernardini13E,Catarina001,Bastos00}.
To see explicitly how to introduce noncommutative effects like an external magnetic field in the harmonic oscillator and free particle systems, it is sufficient to make an analogy between the magnetic field associated frequency and the noncommutative parameters. Thus, for the harmonic oscillator, the analogy is done when one has 
\begin{equation}
\frac{q\, B_0}{2m} \sim \frac{\theta}{2\hbar}m\, \omega_0^2 + \frac{\eta}{2m\, h},
\end{equation}
or, more clearly, the intensity of the magnetic field is given by,
\begin{equation}
B_0 \sim \frac{m^2\, \omega_0^2 \, \theta}{q\, \hbar} + \frac{\eta}{q\, \hbar}.
\end{equation}

From the Hamiltonian for the free particle system, that is, $H = p_i^2/2m$, one notice from the Eq.~(\ref{Eq31}) that the parameter $\theta$ is irrelevant, and the noncommutative effects can be simulated through the analogy,
\begin{equation}
B_0 \sim \frac{\eta}{q\, \hbar}.
\end{equation}

For the GQW in the presence of an external magnetic field, one has to consider the Seiberg-Witten map (see the Appendix) which connects the noncommutative algebra to standard Weyl-Heisenberg algebra of the quantum mechanics \cite{Rosenbaum, Catarina}, as to write the following relation between the noncommutative parameters and the intensity of the magnetic field,
\begin{equation}
B_0 = \frac{\eta}{q \hbar},\quad \mbox{and} \quad r_1 \rightarrow \nu r_1 - \frac{\theta}{2\nu \hbar}p_2 .
\end{equation}
which becomes clearer if an auxiliary parameter is defined by
\begin{equation}
s = \frac{1}{\mu \nu} - 1.
\label{auxliary}
\end{equation}

The equations of motion from (\ref{solutionsLinear}) clearly do not reproduce the solutions as exhibited by (\ref{EMGQW}) when the parameter $\omega$ goes to zero. After turning on the magnetic field, the harmonic feature, so far absent, turns to be much more relevant than the linear term of the Hamiltonian. 
One notices that the two redefined noncommutative parameters, $\theta$ and $\eta$, are constrained by an implicit dependence on $s = s(\mu, \nu)$ and $\omega$, respectively. 
One thus needs to consider that $\eta \neq 0$ in order to ensure that the solutions from (\ref{solutionsLinear}) are not divergent. 
In the very particular case of $\mu\nu \rightarrow 1/2$, that is, for $s \rightarrow 1$, the free particle system in the presence of a magnetic field is recovered by the noncommutative free particle system. On the other hand, when one takes the limit of $\mu\nu \rightarrow 1$, i. e. $s \rightarrow 0$ in Eq.~(\ref{auxliary}), the system exhibits a linear time-dependence, maximizing the intervention of the Seiberg-Witten map represented by the arbitrary parameters $\mu$ and $\nu$.

\section{Conclusions}

The decoherence effects triggered off by external magnetic fields on some elementary quantum mechanical systems were quantified through the quantum fidelity and the Shannon entropy.
The relevant issue here is concerned with the mathematical structure of the Hamiltonian in the harmonic oscillator and free particle systems. It allows one to perform quantitative comparisons with other Hamiltonian structures that exhibit similar dynamics, that is, the noncommutative quantum mechanics. From a point of view of quantum simulation, the results suggest the possibility of building some controllable systems that may exhibit noncommutative profiles.
For instance, the gravitational quantum well turns to be a relevant scenario where one can work on an experimental platform to estimate the values of these noncommutative parameters, $\eta$ and $\theta$ \cite{Bertolami, Banerjee}.
In particular, the study of the noncommutative gravitational quantum well system with a driven laser mechanism detecting the information of the original quantum state has been considered recently (see Ref. \cite{Driven}).
Given the corresponding map in terms of external interacting fields, the phase-space noncomutativity effects can be interestingly considered when investigating the issues related to the interface between quantum and classical descriptions of Nature.

\section*{Appendix: The Seiberg-Witten map as an effective external field}

A theoretical manner to perform a \textit{magnetic field}-like coupling is through the Seiberg-Witten map. This map, in the context of quantum field theory, is due to the connection between the noncommutative quantum mechanics and the standard quantum mechanics \cite{Bernardini01, Rosenbaum}. In fact, the extended noncommutative algebra, which is a case of the modified Heisenberg-Weyl algebra, is represented by the commutation relations given by
\begin{equation}
\left[ \hat{q}_i,  \hat{q}_j \right] = i \theta_{ij} , \hspace{0.5 cm} \left[ \hat{q}_i,  \hat{p}_j \right] = i \hbar \delta_{ij} ,
\hspace{0.5 cm} \left[ \hat{p}_i,  \hat{p}_j \right] = i \eta_{ij} ,  \hspace{0.5 cm} i,j= 1, ... ,d
\label{Eq31}
\end{equation}
where $\eta_{ij}$ and $\theta_{ij}$ are invertible antisymmetric real constant ($d \times d$) matrices, one can define the matrix
\begin{equation}
\Sigma_{ij} \equiv \delta_{ij} + {1\over\hbar^2}  \theta_{ik} \eta_{kj},
\label{Eq32}
\end{equation}
which is an equally invertible if $\theta_{ik}\eta_{kj} \neq -\hbar^2 \delta_{ij}$. 

The Seiberg-Witten map is a linear transformation which maps the algebra from Eq.~(\ref{Eq31}) to the standard commutation relations of the quantum mechanics. This linear transformation ensures that the noncommutative algebra admits a representation in terms of the Hilbert space of ordinary quantum mechanics. Thus, a non-commutative effect in the Hamiltonian of a system works effectively like a external field.

{\em Acknowledgments - The work of AEB is supported by the Brazilian Agencies FAPESP (grant 15/05903-4) and CNPq (grant 300809/2013-1 and grant 440446/2014-7).}

\end{document}